\documentclass[11pt]{article}

\input epsf
\usepackage{epsfig}
\usepackage{latexsym}
\usepackage{amssymb}
\usepackage{amsmath}

\begin{document}
\centerline{\Large \bf Stochastic inflation on the brane}
\vskip 2 cm

\centerline{Kerstin E. Kunze\footnote{E-mail: Kerstin.Kunze@physik.uni-freiburg.de} }

\vskip 0.3cm

\centerline{{\sl Physikalisches Institut,
Albert-Ludwigs-Universit\"at Freiburg, }}
\centerline{{\sl Hermann-Herder-Strasse 3,
D-79104 Freiburg, Germany. }}

\vskip 1.5cm

\centerline{\bf Abstract}
\vskip 0.5cm
\noindent
Chaotic inflation on the brane 
is considered in the context of stochastic inflation.
It is found that there is a regime in which eternal inflation
on the brane takes place. 
The corresponding probability distributions are  
found in certain cases. 
The stationary probability distribution 
over a comoving volume and  
the creation probability of a de Sitter braneworld
yield the same exponential behaviour.
Finally, nonperturbative effects are briefly discussed.

\vskip 1cm

\section{Introduction}
The global structure of many four-dimensional inflationary universes
is very rich \cite{first,llm}. Self-reproduction of inflationary domains leads 
to a universe consisting of many large domains. In each of these
domains there can be different realizations of the inflationary scenario.
This property of standard inflation alleviates the problem of 
fine tuning in inflation. The process of self-reproduction of inflationary domains
leads globally to eternal inflation, namely, there is at least one
inflating region in the universe. 
Self-reproduction of inflationary domains can be understood 
as a branching diffusion process in the space of field values of the inflaton.
The probability distributions found in such a way will give the 
probabilities of finding a certain field value at a certain point in space-time.

There is an interesting connection between the stochastic approach 
to inflation and quantum cosmology. 
The probability of finding the universe in a state
characterized by certain parameters assuming the Hartle-Hawking
no-boundary condition yields the same 
exponential behaviour 
as the stationary probability distribution
in a comoving volume derived from stochastic inflation.
This is the case for standard chaotic inflation. As it turns out 
this also holds for chaotic inflation on the brane.

The idea that the universe is confined to a brane embedded in a higher dimensional
bulk space-time has received much attention in recent years. In the 
one brane Randall-Sundrum  scenario the brane is embedded in
a five-dimensional bulk space-time with negative cosmological constant $\Lambda_5$
\cite{rs,br}.
Chaotic inflation on the brane in this setting has been investigated
in \cite{mwbh,cll,cha}.
The form of the spectrum of perturbations
is modified due to the modified dynamics at high energies.
Therefore, it seems to be interesting to investigate stochastic inflation on the
brane.

\section{Inflation on the brane}
In a braneworld scenario 4D Einstein gravity is recovered on the brane
with some corrections at high energies. Furthermore there are corrections
due to the gravitational interaction with the bulk space-time.
Assuming a fine tuning between the brane tension and the bulk cosmological
constant $\Lambda_5$ leads to the vanishing of the 4D cosmological constant on the
brane. Neglecting contributions from the dark radiation term, this leads
to the following Friedmann equation on the 
brane \cite{mwbh,cll}
\begin{eqnarray}
H^2=\frac{8\pi}{3M_4^2}\rho\left(1+\frac{\rho}{2\lambda}\right),
\end{eqnarray}
where $\rho$ is the energy density on the brane, $\lambda$ the (positive)
brane tension and 
$M_4$ the four-dimensional Planck mass.
This is related to the five-dimensional Planck mass $M_5$
by
\begin{eqnarray}
M_4=\sqrt{\frac{3}{4\pi}}\left(\frac{M_5^2}{\sqrt{\lambda}}\right)M_5.
\end{eqnarray}

Assuming that matter on the brane is dominated by a scalar field $\phi$,
confined to the brane, with potential $V(\phi)$, its 
equation of motion is given by
\begin{eqnarray}
\ddot{\phi}+3H\dot{\phi}+V'(\phi)=0.
\end{eqnarray}
In the slow roll approximation,
\begin{eqnarray}
H^2&\simeq&\frac{8\pi}{3M_4^2}V\left(1+\frac{V}{2\lambda}\right)
\label{fri-ex}\\
\dot{\phi}&\simeq&-\frac{V'}{3H}.
\end{eqnarray}
For $\lambda\rightarrow\infty$ the usual Friedmann equation is recovered.
For $V\gg\lambda$  brane effects dominate.

Inflation takes place if the Hubble parameter satisfies
$|\dot{H}|<H^2$.
In a braneworld with matter given by a scalar field the 
condition for inflation yields to \cite{mwbh}
\begin{eqnarray}
\dot{\phi}^2-V+\frac{\dot{\phi}^2+2V}{8\lambda}\left(5\dot{\phi}^2-2V\right)<0.
\label{con1}
\end{eqnarray}

In the following, the potential of the inflaton will be taken to be of the
form,
\begin{eqnarray}
V=\frac{1}{2}m^2\phi^2.
\end{eqnarray}
Furthermore, for convenience, everything will be expressed  
in four-dimensional Planck units, hence $M_4\equiv 1$.

\section{Eternal inflation on the brane}

The evolution of a scalar field in an inflationary universe 
is determined by two contributions. On the one hand, there is 
the classical rolling down of the scalar field down its 
potential. On the other hand, there are quantum fluctuations
of the inflaton which become classical outside the horizon.
This latter contribution can have either positive or negative sign.
The classical rolling down is given by,
$\Delta\phi\simeq\dot{\phi}\Delta t$, where 
in the slow roll approximation $\dot{\phi}$ is given by
$\dot{\phi}\simeq-\frac{V'}{3H}$.
The amplitude of a quantum fluctuation is given
by $\delta\phi=\frac{H}{2\pi}$. In a typical time 
interval $H^{-1}$, 
$e^3$ new domains appear each containing an almost 
homogeneous field $\phi-\Delta\phi+\delta\phi$ \cite{llm}.
There is a critical value $\phi_s$ for which for all $\phi\geq \phi_s$
quantum fluctuations dominate over the classical
evolution towards smaller field values.
This is the regime of self-reproduction of inflationary 
domains.
$\phi_s$ is determined by \cite{llm96}
\begin{eqnarray}
\frac{2\pi}{3}\left.\frac{V'}{H^3}\right|_{\phi=\phi_s}=1.
\label{3.7}
\end{eqnarray}
For field values $\phi\geq\phi_s$ there will be domains in which quantum jumps 
lead to an increase in the field value of the inflaton. In a 
small percentage of domains this will lead to the 
maximal field value at which inflation takes place.
The upper boundary $\phi_{5{\rm D}}$ in this braneworld model is determined by
the five-dimensional Planck boundary.
For energies higher than $V(\phi_{5{\rm D}})=M_5^4$
the scalar field becomes deconfined and flows off the brane
into the bulk \cite{cll}.
As in the four-dimensional case inflation will stop 
at this boundary. Here one might argue that inflation
on the brane stops since the scalar field is flowing off the
brane and thus four-dimensional inflation can no longer be sustained
on the brane.
Thus the upper boundary is given by 
\begin{eqnarray}
\phi_{5{\rm D}}=\sqrt{2}
\left(\frac{4\pi}{3}\right)^{\frac{1}{3}}
\frac{\lambda^{\frac{1}{3}}}{m}.
\end{eqnarray}

In the low energy regime, $V\ll\lambda$ the Friedmann
equation on the brane reduces to the standard one,
\begin{eqnarray}
H^2\simeq\frac{8\pi}{3} V.
\label{fri-lo}
\end{eqnarray}
It will be assumed that the whole period of inflation takes place in the 
low energy regime.
The end of inflation $\phi_e$ is determined by the first two terms 
in (\ref{con1}), $\dot{\phi}^2=V$, which yields,
$\phi_e=1/\sqrt{6\pi}$.
Furthermore, $\phi_e<\phi_{5{\rm D}}$ implies
$\lambda>(12\pi)^{-3/2}(3/4\pi)m^3$.
The lower boundary for self-reproduction  $\phi_s$ is
given by 
$$\phi_s=\left(\frac{3}{16\pi}\right)^{\frac{1}{4}}m^{-\frac{1}{2}}.$$
The requirement $V/2\lambda<1$ at $\phi_{5{\rm D}}$ implies
$\lambda>2\pi^2/9$. 
Eternal inflation takes place if
$\phi_s<\phi_{5{\rm D}}$ which yields the condition, 
$\lambda>(1/8)(3/4\pi)^{7/4}m^{3/2}$.
In the standard inflationary scenario observations
require that $m\simeq 10^{-13} {\rm GeV}=10^{-6} M_4$.
Thus for realistic values of $m$ eternal inflation takes place 
on the brane in the low energy regime, as long as $\lambda>2\pi^2/9$.
However, note that inflation as well as self-reproduction 
takes place at field values larger than the four-dimensional
Planck scale.

In the high energy limit, $V\gg\lambda$, the 
quadratic term in the Friedmann equation dominates,
\begin{eqnarray}
H^2\simeq\frac{8\pi}{3}\frac{V^2}{2\lambda}.
\label{fri-hi}
\end{eqnarray}
Inflation is assumed only to take place in the 
high energy regime.
The end of inflation is determined by the last two terms 
in the condition (\ref{con1}), 
$5\dot{\phi}^2=2V$, implying,
$\phi_e=(5/3\pi)^{1/4}\lambda^{1/4}m^{-1/2}$.
The lower boundary for eternal inflation $\phi_s$ is given by
$$\phi_s=\left(\frac{12}{\pi}\right)^{\frac{1}{10}}
\lambda^{\frac{3}{10}}m^{-\frac{4}{5}}.$$
Out of the two inequalities $\phi_e<\phi_{5{\rm D}}$ and 
$\phi_s<\phi_{5{\rm D}}$ it is found that 
the first one provides the stronger bound on the brane 
tension $\lambda$, namely, $\lambda>8\times 10^{-6} m^6$.
An upper bound on $\lambda$ is found by requiring that 
$V(\phi_e)/2\lambda>1$, implying $\lambda <3\times 10^{-3}m^2$.
As shown in \cite{mwbh} in the limit of strong brane
corrections the COBE normalization of the curvature perturbations
put a bound on $m$, which can be written in units of $M_4$ as,
$m\sim 6\times 10^{-5}\lambda^{1/6}$. 
This implies $\phi_e\sim 10^2\lambda^{1/6}$,
$\phi_s\sim 3\times 10^3 \lambda^{1/6}$,
and $\phi_{5{\rm D}}\sim 4\times 10^4 \lambda^{1/6}$.
Thus $\phi_e<\phi_s<\phi_{5{\rm D}}$ and eternal inflation 
takes place for values of $m$ derived from observations.
As already pointed out in \cite{mwbh},
the 5D Planck boundary $\phi_{5{\rm D}}$ is 
below the 4D Planck scale, $M_{4}$. Thus eternal
inflation takes place at field values below $M_{4}$.

For a certain range of parameters eternal inflation
takes place inside the low energy and inside
the high energy regime on the brane.
In this case domains reproduce themselves.
Since it was assumed that 
the inflationary period is either in the low or 
in the high energy regime
the dynamics of each of them will be 
determined by the 
characteristics of the regime that they are
originating from.
However, it could also be considered that a 
domain starts in the low energy regime and 
then due to the process of stochastic inflation,
field values in successive domains reach 
such high values that
strong brane corrections become important.
Thus in this case the Friedmann equation on the
brane changes from equation (\ref{fri-lo})
to equation (\ref{fri-hi}).
In order for this to happen, one has to require that 
eternal inflation takes place in the low energy regime.
Furthermore, the 5D Planck boundary has to be 
in the high energy regime. This implies that the
brane tension has to satisfy, $\lambda<2\pi^2/9$.
In this case, out of a low energy domain domains 
with the low energy characterics emerge as
well as those with the high energy dynamics emerge.
This picture is similar to the model proposed in \cite{lz}
where the space-time dimension can change locally in chaotic
eternal inflation. In this case, on the brane there are 
regions in which the dynamics are  determined by the high 
energy Friedmann equation (\ref{fri-hi})
and there are domains in which the Friedmann equation
is the low energy one (\ref{fri-lo}).

\section{Stochastic description}
The stochastic nature of the effect of the competition between the 
classical rolling down and the quantum perturbations is
captured by a Fokker-Planck equation \cite{fok}.
The (classical) field $\phi$ is performing a Brownian motion
described by a Langevin equation, \cite{first,llm}
\begin{eqnarray}
\frac{d}{dt}\phi=-\frac{V'(\phi)}{3H(\phi)}+\frac{H^{\frac{3}{2}}(\phi)}{2\pi}
\xi(t),
\end{eqnarray}
where $\xi(t)$ describes the white noise due to the quantum fluctuations,
which causes the Brownian motion of the classical field $\phi$.

The probability distribution $P_c(\phi,t)$ 
determines the probability to 
find a given value of the field $\phi$ at a given  time at a
given point. This is the probability distribution over a
comoving coordinate volume, i.e. over a physical volume
at some initial moment of inflation.
$P_c(\phi,t)$ is determined by the equation \cite{llm},
\begin{eqnarray}
\frac{\partial P_c}{\partial t}=
\frac{\partial}{\partial\phi}\left[
\frac{H^{3(1-\beta)}(\phi)}{8\pi^2}\frac{\partial}{\partial\phi}
\left(H^{3\beta}(\phi)P_c\right)+
\frac{V'}{3H(\phi)}P_c
\right].
\end{eqnarray}
The parameter $\beta$ encodes an ambiguity in the derivation of this
equation for systems for which the diffusion coefficient depends on
$\phi$. $\beta=1$ corresponds to the It\^o version of stochastic 
analysis and $\beta=\frac{1}{2}$ to the Stratonovich version \cite{fok}.
An exact  stationary solution, for which $\frac{\partial P_c}{\partial t}=0$,
can be found for the Hubble parameter given by equation 
(\ref{fri-ex}), namely,
\begin{eqnarray}
P_c(\phi)=\left(\frac{8\pi}{3}\right)^{-\frac{3\beta}{2}}
V^{-\frac{3\beta}{2}}\left[1+\frac{V}{2\lambda}
\right]^{-\frac{3\beta}{2}}
\left[1+\frac{2\lambda}{V}\right]^{-\frac{3}{8\lambda}}
\exp\left[\frac{3}{8V}\;
\frac{1+\frac{V}{\lambda}}{1+\frac{V}{2\lambda}}
\right].
\label{pc}
\end{eqnarray}
In the low energy regime, $V\ll\lambda$, this reduces to the well known
expression  \cite{llm}
\begin{eqnarray}
P_c\sim V^{-\frac{3\beta}{2}}(\phi)\exp\left(\frac{3}{8V(\phi)}
\right).
\label{lep}
\end{eqnarray} 
Considering the high energy regime, $V\gg\lambda$, the stationary distribution $P_c(\phi)$
approaches,
\begin{eqnarray}
P_c(\phi)\simeq\left(\frac{4\pi}{3}\right)^{-\frac{3\beta}{2}}
\left(\frac{V^2}
{\lambda}\right)^{-\frac{3\beta}{2}}
\exp\left[
\frac{\lambda^2}
{2V^3}\right].
\label{p_c_high}
\end{eqnarray}
It is interesting to compare the expression for the stationary 
probability distribution (\ref{pc})
with the probability for creation of a braneworld from
nothing.
This is described by the de Sitter brane instanton \cite{gs}.
The probability for creation of a universe in the 
Hartle-Hawking no-boundary proposal \cite{HH}
is given by, ${\cal P}\sim \exp(-S_{E})$, with $S_E$ the 
Euclidean action.
In the case of the creation of a braneworld containing
just one de Sitter brane in an AdS bulk, the Euclidean action, 
in the notation used here,
is given by \cite{gs},
\begin{eqnarray}
S_E=-\frac{\pi}{H^2}\left(1+\frac{V}{\lambda}\right).
\end{eqnarray}
Using the expression for the Hubble parameter 
on the brane (\ref{fri-ex}) the nucleation probability of
a de Sitter braneworld is given by
\begin{eqnarray}
{\cal P}\sim\exp\left(\frac{3}{8V}\;
\frac{1+\frac{V}{\lambda}}{1+\frac{V}{2\lambda}}\right).
\label{p}
\end{eqnarray}
Thus comparing the exponentials in the stationary probability
distribution $P_c$ (\ref{pc}) and in the 
probability distribution ${\cal P}$ (\ref{p})
it is found that they are exactly the same. 
Therefore the same coincidence between these two
probability distributions appears as is the case in
standard four-dimensional inflation \cite{llm}.

$P_c(\phi,t)$ is the probability distribution 
in a comoving volume, neglecting the expansion of the universe.
The probability distribution $P_p(\phi,t)$ in a proper volume takes into
account that during a small time interval $dt$ the total number of 
points associated with the field $\phi$ is additionally increased
by a factor $3H(\phi)dt$. Thus this leads
to the following equation \cite{llm96,first}
\begin{eqnarray}
\frac{\partial P_p}{\partial t}=
\frac{\partial}{\partial\phi}\left[
\frac{H^{3(1-\beta)}(\phi)}{8\pi^2}\frac{\partial}{\partial\phi}
\left(H^{3\beta}(\phi)P_p\right)+
\frac{V'}{3H(\phi)}P_p
\right]+3H(\phi)P_p(\phi,t).
\label{p_p}
\end{eqnarray}
In order to solve this equation it is convenient to make the ansatz \cite{llm}
\begin{eqnarray}
P_p(\phi,t)=\sum_{s=1}^{\infty}e^{\lambda_s t}\pi_s(\phi)
\sim
e^{\lambda_1 t}\pi_1(\phi)
\;\;\;\;\;\;\;\;\;\;\;\;\;\;\;\;
{\rm for}\;\;{t\rightarrow\infty}
\label{an}
\end{eqnarray}
For large times $t\rightarrow\infty$ only the 
largest eigenvalue is kept.
In the high energy regime brane effects are dominant and the deviations
from standard 4D inflation are the largest. Therefore, in the following,
the probability distribution $P_p$ will be discussed for an
inflationary period entirely in the high energy regime.
Thus the Hubble parameter is given by (\ref{fri-hi}). 
Together with this and (\ref{an})
equation (\ref{p_p}) yields to,
\begin{eqnarray}
\pi_1''+\left[\frac{9}{\phi}+24\frac{\lambda^2}{m^6}\frac{1}{\phi^7}
\right]\pi_1'
+\left[\frac{15}{\phi^2}+72\pi\frac{\lambda}{m^4}\frac{1}{\phi^4}
-8\pi^2\left(\frac{\pi}{3}\right)^{-\frac{3}{2}}
\lambda_1 
\frac{\lambda^{\frac{3}{2}}}{m^6}\frac{1}{\phi^6}
-24\frac{\lambda^2}{m^6}\frac{1}{\phi^8}\right]\pi_1=0.
\end{eqnarray}
where $\beta=\frac{1}{2}$.
The boundary conditions on $P_p$ are equivalent to those 
in the standard four-dimensional case \cite{llm}.
There is no diffusion below the end of inflation, which implies
\newline
$\frac{\partial}{\partial\phi}\left(H^{\frac{3}{2}}(\phi)P_p
\right)_{\phi_e}=0$, and inflation stops
at the 
(5D) Planck boundary,
thus
$P_p(\phi_{5{\rm D}})=0$. This imposes the following conditions on 
$\pi_1(\phi)$,
\begin{eqnarray}
\pi_1'(\phi_e)=-\frac{3}{\phi_e}\pi_1(\phi_e)\hspace*{2cm}
\pi_1(\phi_{5{\rm D}})=0.
\end{eqnarray}
Numerical solutions 
have been plotted in figure 1.
\begin{figure}
\centerline{\epsfxsize=1.8in\epsfbox{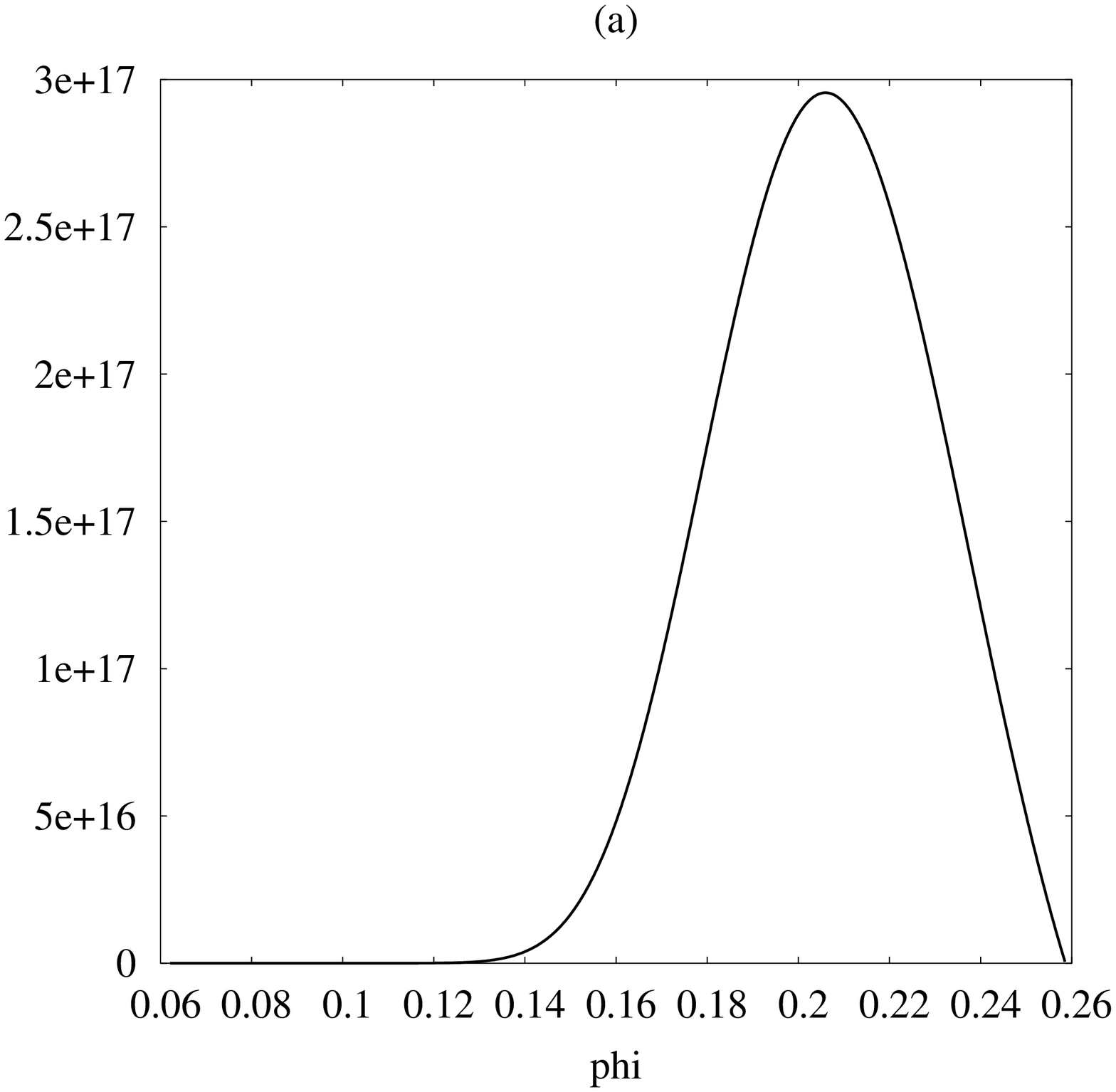} \hspace{0.1cm}
              \epsfxsize=1.8in\epsfbox{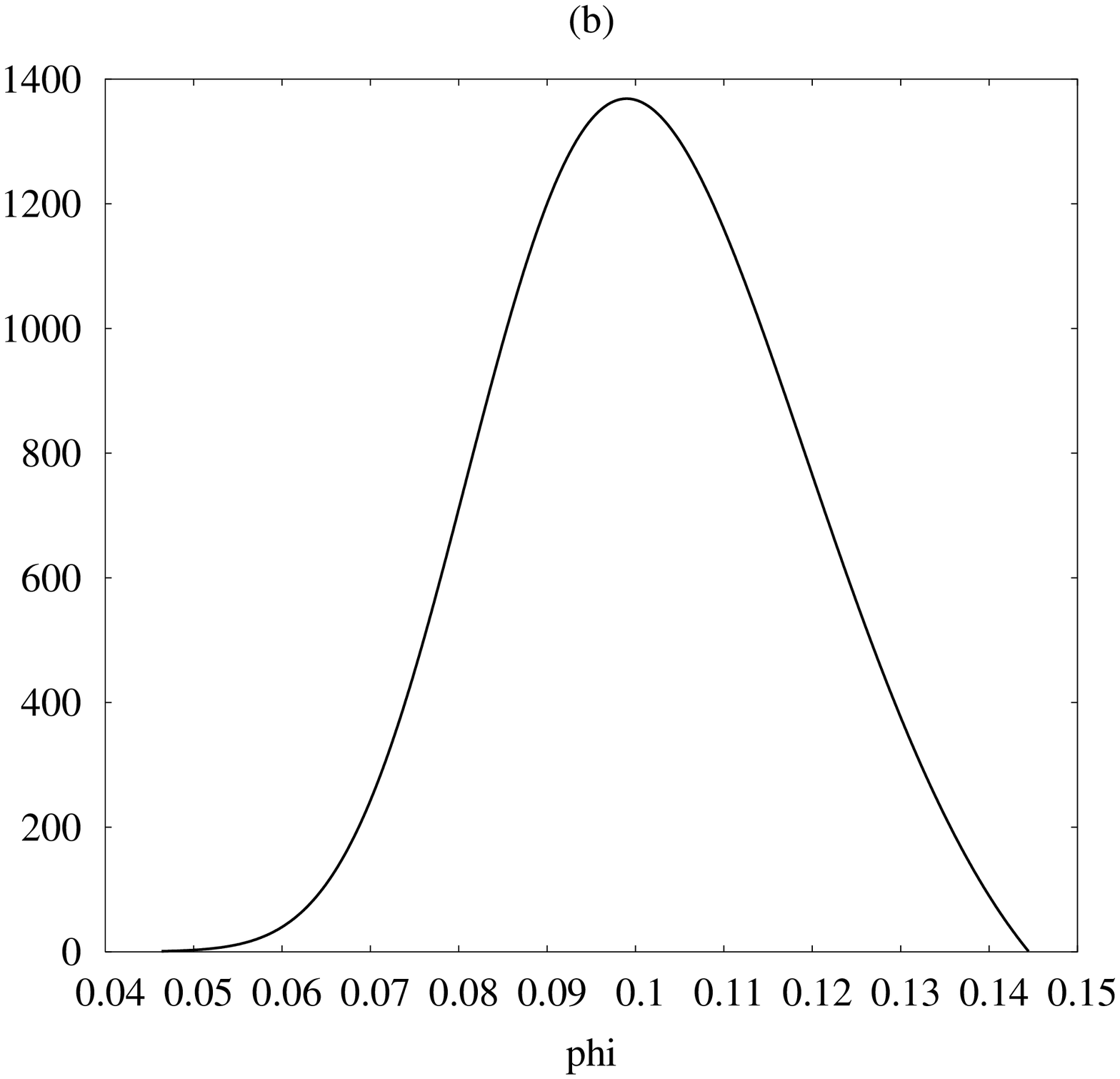}
\hspace{0.1cm}
              \epsfxsize=1.8in\epsfbox{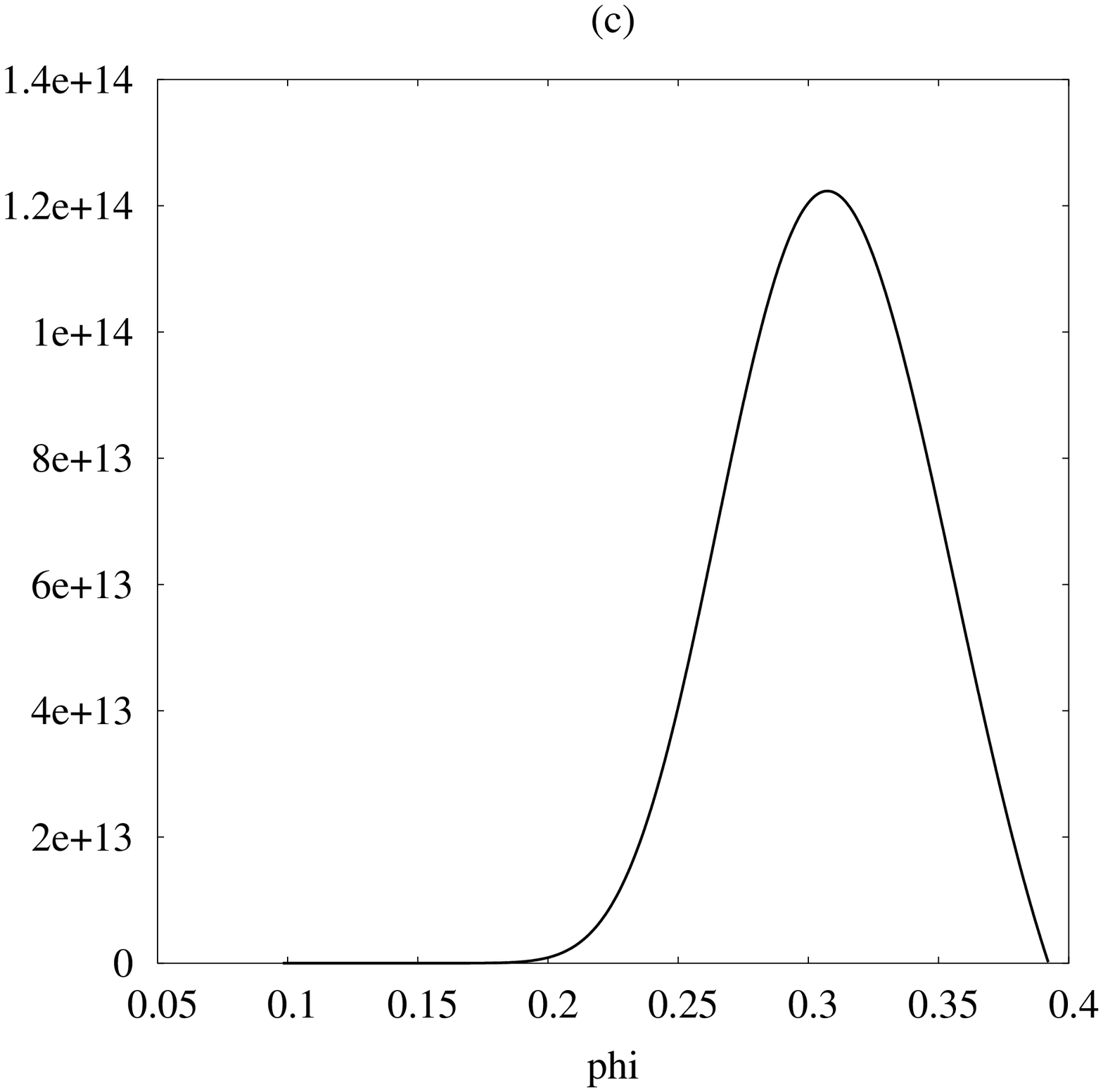}}
\caption{(a) shows the numerical solution for $\pi_1(\phi)$
for $\lambda=10^{-8}$, $m=1.9\times10^{-2}$, $\lambda_1=0.417$.
(b) shows $\pi_1(\phi)$ for $\lambda=10^{-8}$, 
$m=3.4\times10^{-2}$, $\lambda_1=0.316$.
(c) shows $\pi_1(\phi)$ for
$\lambda=2\times10^{-7}$, $m=3.4\times10^{-2}$,
$\lambda_1=0.666$.}
\end{figure}
In figure 1, the probability 
distribution shows a maximum in all cases. 
For larger values of $m$ at constant brane tension $\lambda$
it is shifted towards smaller field values $\phi$.
For larger values of $\lambda$ at constant $m$ it is shifted 
towards higher field values $\phi$, concentrated very close
to the 5D Planck boundary. 
The eigenvalue $\lambda_1$ can be related to the fractal dimension
of the universe, $d_{fr}$, defined as  $d_{fr}=\lambda_1/H_{max}$ 
\cite{llm,vil},
where $H_{max}$ is the maximal Hubble parameter at the five-dimensional
Planck boundary. The fractal dimension is motivated by the observation that 
at the Planck boundary the total volume of inflationary domains
does not grow as $e^3$ during a time interval $H_{max}^{-1}$
but only as $\exp(\lambda_1/H_{max})$. Some domains will 
reach energies beyond the Planck scale and thus drop out of
the total volume.
On the brane in the high energy regime the maximal 
Hubble parameter at the 5D Planck boundary is given by
$H_{max}=\left(\frac{4\pi}{3}\right)^{\frac{7}{6}}\lambda^{\frac{1}{6}}$.
Thus the numerical examples in figure 1 have fractal dimensions
much below 3.

For scalar field values $\phi>\phi_s$ self-reproduction takes 
place. This phenomenon is related to the quantum jumps that cause 
the field values to increase. 
However, as pointed out in \cite{llm96} these quantum jumps could also 
coherently add up to lead to a larger than usual jump down the 
potential. This might lead to nonperturbative amplification of
inhomogeneities. 
Domains in which quantum jumps lead to an increase
of the field value are pushed up to the five dimensional
Planck boundary. At this point the Hubble parameter 
reaches its maximum, $H_{max}$. 
Thus these domains give the greatest contribution to the volume 
of the universe. The domains will stay as long as possible at the 
Planck boundary and then ''rush down'' the potential
with an amplitude larger than $H/2\pi$ \cite{llm96}.
Following \cite{llm96} the extra time, $\tilde{\Delta} t$ spent at the 
Planck boundary can be estimated by,
$\tilde{\Delta} t(\phi)=\tilde{\delta}\phi/\dot{\phi}$,
where $\tilde{\delta}\phi$ is the amplified amplitude of the 
jump down the potential, 
$
\tilde{\delta}\phi=n(\phi)H(\phi)/(2\pi),
$
where $n(\phi)$ is an amplification factor.
In the regime where brane effects are dominant, $V\gg\lambda$, this leads to
\begin{eqnarray}
\tilde{\Delta} t(\phi)=2n(\phi)\frac{V^2}{\lambda V'}.
\end{eqnarray}
The volume is increased by a factor $\exp(d_{fr} H_{max} \tilde{\Delta}t )$.
Hence the volume-weighted probability (\ref{an}) is given by \cite{llm96}
\begin{eqnarray}
P\sim\exp\left[d_{fr} H_{max} \tilde{\Delta} t(\phi)-\frac{1}{2}n^2(\phi)\right],
\end{eqnarray}
where it is assumed that an amplification of the standard jump is 
suppressed by a factor $\exp[-\frac{1}{2}n^2(\phi)]$.
Maximizing this with respect to $n(\phi)$ gives 
\begin{eqnarray}
n(\phi)=2d_{fr}H_{max}\frac{V^2}
{\lambda V'}.
\end{eqnarray}
Expressing this in terms of the ratio 
of the amplitudes of scalar to tensor perturbations,
$A_S/A_T$ \cite{lmw}, yields to
\begin{eqnarray}
n(\phi)=\left(\frac{3}{2\pi}\right)^{\frac{1}{2}}
d_{fr}H_{max}
\left(\frac{V}{\lambda}\right)^{\frac{1}{2}}
\frac{A_S}{A_T}.
\end{eqnarray}
In the standard four-dimensional case 
dependence on the inflaton field $\phi$ in 
the amplifying factor $n(\phi)$ can be entirely expressed in terms of
the ratio of the amplitudes of the scalar to tensor perturbations.
As it turns out, in the high energy regime of chaotic inflation
on the brane this is no longer the case. There is an additional
amplifying factor $V/\lambda$.
Since $A_S/A_T$ is an observable quantity this means that
the amplitude of the jumps down the 
potential are amplified with respect to the case of 
standard four-dimensional inflation.
Domains which jump down with these amplified 
amplitudes end up as regions with smaller energy 
density compared to the background. In a braneworld 
these wells or infloids \cite{llm96}
are deeper than in standard four-dimensional inflation.

\section{Conclusions}
The stochastic approach to standard 4D inflation 
and its variations opened 
the way to a rich global structure of an inflating 
universe. Here the stochastic approach to inflation has been 
applied to a braneworld model, namely, to chaotic inflation
on the brane. It has been shown that eternal inflation
takes place for a certain range of parameters, and in particular,
for those satisfying observational bounds.

The competition between the evolution towards smaller field values due
to classical dynamics and the evolution towards either even smaller or higher 
field values can be described as a Brownian motion.
There exists a well defined procedure to obtain 
a Fokker-Planck equation determining the probability 
distribution to find a certain value of the scalar field at
a given point in space-time. Furthermore, there are two types 
of probability distributions. Firstly, the probability distribution 
$P_c$ in a given comoving volume. Secondly, the probability
distribution $P_p$ in a given physical volume, which takes into account
the expansion of the universe.
In standard 4D chaotic inflation, apart from some pre-factors, the 
dominant behaviour is determined by an exponential function, which
is exactly the square of the Hartle-Hawking no-boundary wavefunction
of the universe. In the braneworld scenario discussed here, a similar result 
was found.
Comparing the expression for $P_c$ found in the stochastic approach
to chaotic inflation on the brane with the de Sitter brane instanton 
for a one brane system as calculated in \cite{gs},
apart from some prefactors, the same exponential
function was found in the two cases.

The probability distribution in a given physical 
volume, $P_p$, was calculated numerically in the high energy regime
where brane effects dominate. The results are similar to the 
ones in standard 4D inflation with the distribution
concentrated near the 5D Plack boundary. 

Finally, the process of a scalar field close to the 5D Planck
boundary rolling down with amplitudes larger than the usual
$H/2\pi$, due to quantum fluctuations, was briefly discussed.
It was found that the amplification factor is enhanced 
by a factor $V/\lambda$ in the high energy regime 
on the brane. Thus the infloids or wells in the energy
distribution are deeper than in  standard four dimensional
inflation.

\section{Acknowledgments}
It is a pleasure to thank J. Garriga and M.A. V\'azquez-Mozo for 
enlightening discussions.
I would like to thank the University of Geneva for hospitality 
where part of this work was done.
This work has been supported in part by Spanish Science Ministry
Grant FPA 2002-02037.

\end{document}